\documentclass[twoside]{article}
\usepackage{fleqn,myproc,amssymb,psfig,epsfig}

\usepackage{graphicx}
\usepackage[figuresright]{rotating}

\newcommand{\AmS}{{\protect\the\textfont2
  A\kern-.1667em\lower.5ex\hbox{M}\kern-.125emS}}
\def\shat{\hat{s}}

\def\sp{{\mathbf s}_1}
\newcommand{\sm}{{\mathbf s_2}}
\newcommand{\kh}{{\hat{\mathbf k}}}
\newcommand{\ph}{\hat{\mathbf p}}
\newcommand{\dhh}{\hat{\mathbf d}}
\title{QCD corrections to top quark spin correlations at hadron colliders
\thanks{Report-no. DESY 00-120}}
\author{W. Bernreuther\thanks{Supported by BMBF 
contract 05 HT9 PAA 1.}\address[MCSD]{Institut f.\ Theoretische Physik, 
RWTH Aachen, \\ 
      D-52056 Aachen, Germany}%
        , A. Brandenburg\thanks{Speaker at the conference. 
Supported by a Heisenberg fellowship of D.F.G.}\address{DESY 
Theory Group,
 D-22603 Hamburg, Germany}, and Z.G. Si\thanks{Supported by a A. v. Humboldt
fellowship.}\addressmark[MCSD]}
\begin{document}
\begin{abstract}
Future hadron collider experiments will allow for a detailed investigation
of $t\bar{t}$ spin correlation effects. In this talk, recent progress
in the theoretical description of these effects is reported. 
In particular, next-to-leading order results for the $t\bar t$ spin
correlations in $q\bar{q}\to t\bar{t}X$ are presented, 
using various spin quantization axes.
\vspace{1pc}
\end{abstract}

% typeset front matter (including abstract)
\maketitle
Top quark pair production and decay can be 
described by perturbative methods. This is so because the top
quark width is large compared to the QCD hadronisation scale and therefore
the top decays before hadronisation takes place.
The full information of the top quark interactions including spin
correlations between $t$ and $\bar{t}$ can thus in principle be 
obtained from studying  differential distributions of the decay
products. In this talk, we will report on recent progress
in the theoretical description of top quark spin correlation
effects at hadron colliders. 

Consider the reactions
\begin{eqnarray}\label{eq:hadronreac}
p\bar{p},pp\to t\bar{t}X\to
\left\{\begin{array}{c}
2\ell+n  \ge 2\ {\mbox{jets}}+P_T^{\rm miss} \\ 
\ell+n  \ge 4\ {\mbox{jets}}+P_T^{\rm miss} \\
n  \ge 6\ {\mbox{jets}}.
 \end{array} \right. 
\end{eqnarray}
To leading order in the QCD coupling, two parton reactions
contribute:
\begin{eqnarray}\label{eq:partonreac}
q\bar{q},gg\to t\bar{t} \to bW^+\bar{b}W^-\to 6 {\mbox{ fermions}}.
\end{eqnarray}
The amplitudes for these two- to six-body
processes, with intermediate top quarks of non-zero total width, were
given first in \cite{Kleiss:1988}.  
The calculation of the fully differential 
cross sections for  the reactions (\ref{eq:hadronreac}) including NLO QCD
corrections is a formidable task. 
A substantial simplification in the computation of the radiative corrections
is achieved if one performs an expansion in 
$\Gamma_t/m_t$ and $\Gamma_W/m_W$ and keeps only the terms with the highest 
degree of resonance. In this leading pole approximation 
\cite{Stuart:1991} the radiative corrections 
can be classified into so-called factorisable and non-factorisable 
corrections. The non-factorisable NLO QCD corrections were calculated in
\cite{Beenakker:1999}. For the factorisable corrections
the square of the complete
 matrix element ${\cal M}^{(\lambda)}$ for the partonic 
reactions (\ref{eq:partonreac}) is
of the form
\begin{equation}
\mid {\cal M}^{(\lambda)}{\mid}^2 \propto {\rm Tr}\;[\rho
R^{(\lambda)}{\bar{\rho}}]
 = \rho_{\alpha'\alpha}
R^{(\lambda)}_{\alpha\alpha',\beta\beta'}{\bar{\rho}}_{\beta'\beta} .
\label{eq:trace}
\end{equation}
Here $R^{(\lambda)}$ denotes the density matrix for the production of on-shell
$t\bar t$ pairs, the label $\lambda$ indicates the process, and
$\rho,{\bar{\rho}}$ are the density matrices describing the decay
of polarised $t$ and $\bar t$ quarks, respectively, into specific final states.
The subscripts in  (\ref{eq:trace}) denote the  $t$, $\bar t$ spin indices.
Note that both the production and decay density matrices are gauge invariant.
\par
For the process $q\bar{q}\to t\bar{t}$, which is the dominant process
at the Tevatron, 
the production density matrix $R$ is given 
in terms of transition matrix elements as follows:
\begin{eqnarray}
R_{\alpha\alpha' ,\beta\beta'}=\!\!
\sum_{{{\rm\scriptscriptstyle colors} \atop
{\rm\scriptscriptstyle q\bar{q}}\;
{\rm\scriptscriptstyle spins} }}\!
\frac{\langle t_\alpha\bar t_\beta |{\cal T}|
\,q\bar q\,\rangle
\langle \,q \bar q\,|{\cal T}^\dagger|
t_{\alpha'}\bar t_{\beta'}\rangle}{N_{q\bar{q}}},
\label{eq:Rdef}
\end{eqnarray}
where
the factor $N_{q\bar{q}} = (2N_C)^2=36$ averages over the spins and colors
of the initial $q\bar q$ pair. 
The matrix structure of $R$  is
\newpage
\begin{eqnarray}
R_{\alpha\alpha',\beta\beta'}&=&
A \delta_{\alpha\alpha'}\delta_{\beta\beta'}+
C_{ij}(\sigma^i)_{\alpha\alpha'}
(\sigma^j)_{\beta\beta'}  \nonumber \\
&+&B_{i} (\sigma^i)_{\alpha\alpha'} 
\delta_{\beta\beta'}
+{\bar B}_{i} \delta_{\alpha\alpha'}
(\sigma^i)_{\beta\beta'} \ ,
\label{eq:Rstruct}
\end{eqnarray} 
\par\noindent
where $\sigma^i$ are the Pauli matrices. Using rotational invariance the
`structure functions'  $B_i,{\bar B}_i$ and
$C_{ij}$ can be
further decomposed.   The function
$A$, which determines the ${\bar t}t$ cross section, is known
to next-to-leading order in $\alpha_s$ from the work of
\cite{Nason:1988,Beenakker:1991}. The corresponding function
for the gluon fusion process is also known in NLO 
\cite{Nason:1988,Beenakker:1989}. Because
of parity (P) invariance  the vectors  ${\bf B},{\bf\bar B}$ can have,
within QCD,  only a component normal to the scattering plane. This component,
which amounts to a normal polarisation of the $t$  and $\bar t$ quarks,
 is induced by the absorptive part of the scattering amplitude,
and it was computed for $q\bar q$ and $g g$ initial states in
\cite{Bernreuther:1996,Dharmaratna:1996}
to order $\alpha^3_s$.  The normal polarisation is quite small, both for
$t\bar t$ production
at the Tevatron and at the LHC. Parity and CP invariance of QCD dictates that
the functions $C_{ij}$, which encode the  correlation  between the  $t$ and
${\bar t}$ spins,
have the structure \cite{Bernreuther:1994}
\begin{equation}
 C_{ij} = c_1\delta_{ij} + c_2
{\hat p}_{i}{\hat p}_{j} + c_3
{\hat k}_{i}{\hat k}_{j} + c_4
({\hat k}_{i}{\hat p}_{j} + {\hat p}_{i}{\hat k}_{j}) ,
\label{eq:cij}
\end{equation}
where ${\ph}$ and ${\kh}$ are the directions of flight of the
initial quark  and of the $t$ quark, respectively, in
the parton c.m. frame. The production density matrix for the reaction
$gg\to t\bar{t}$ can be defined and decomposed in an analogous fashion.
To Born approximation the  functions $c_r$ were given, e.g., in
\cite{Brandenburg:1996}.
Theoretical studies of spin correlations
have been performed at leading order in $\alpha_s$ in 
\cite{Brandenburg:1996,Barger:1989,Stelzer:1996,Mahlon:1996,Mahlon:1997,Chang:1996}. 
An attempt to detect spin correlations in a small $t\bar t$ dilepton
sample collected at the Tevatron was recently reported by the D0 collaboration
\cite{D0:2000}. A feasibility study for the LHC can be found in
\cite{Top:2000}. 

We have recently computed the $t\bar t$ spin
density matrices
for the parton reactions $q {\bar q} \to t{\bar t}, t{\bar t} g$ to order
$\alpha_s^3$ \cite{Bernreuther:2000}.
We have also calculated, for these reactions, the 
degree of the ${t \bar t}$ spin correlation
at NLO for different $t$ and $\bar t$ spin quantization axes.
\par
The $t\bar t$
spin correlations 
can be inferred from appropriate angular correlations and distributions
of the $t$ and $\bar t$  decay products. In the SM the main top decay modes are
$t\to b W \to b q {\bar q}',  b \ell \nu_{\ell}$. Among these final states the
charged leptons, or the jets from quarks of weak isospin -1/2 originating
from $W$ decay, are most sensitive to
the polarisation of the top quarks.
The one-loop QCD corrections to the semileptonic decays of polarised top
quarks and to
$t \to W + b$ can be extracted from the results of \cite{Czarnecki:1991} and
\cite{Schmidt:1996,Fischer:1999}, respectively. In the following we 
describe our
computation of the  density matrices for $t\bar t$ production by $q\bar q$
annihilation.
At NLO we have to consider the reactions
\begin{equation}
q(p_1) + {\bar q}(p_2) \rightarrow t(k_1) + {\bar t}(k_2),
\label{eq:qq}
\end{equation}
and
\begin{equation}
q(p_1) + {\bar q}(p_2) \rightarrow t(k_1) + {\bar t}(k_2) + g(k_3).
\label{eq:qqgluon}
\end{equation}
In order to determine these functions to
order $\alpha_s^3$  we first computed the one-loop diagrams that contribute to
(\ref{eq:Rdef}). Dimensional regularization was employed to treat both the
ultraviolet
and the infrared and collinear singularities which appear in the diagrams.
The ultraviolet singularities were removed by
using the $\overline{\rm{MS}}$ prescription for the QCD coupling $\alpha_s$
and the on-shell definition of the top mass $m_t$. The initial quarks are
taken to be
massless. After renormalisation the density matrix for the $t\bar t$ final
state still contains single and double poles in $\epsilon = (4-D)/2$ 
due to soft and collinear singularities.
These poles are cancelled after including the contributions of the reaction
(\ref{eq:qqgluon}) and  mass factorization. For the latter we
used the $\overline{\rm{MS}}$ factorization scheme. 
We avoided the computation of the exact density matrix for
the reaction (\ref{eq:qqgluon}) in $D$ dimensions by employing a
simple version of the phase-space slicing
method \cite{Giele:1993}: We divided the phase space
into four regions, namely the region where the gluon is soft, the two 
regions where the gluon is  
collinear to one of the initial state massless
quarks (but not soft), and the complement of these three regions, where
all partons are `resolved'.
This decomposition can be performed using a single dimensionless 
cut parameter $x_{\rm min}$. For example, the soft region is defined
by the requirement that the scaled gluon energy
in the c.m. system $x_g=2E_g/\sqrt{s}$ is smaller than $x_{\rm min}$.
 In the soft region we used the eikonal
approximation of the matrix element for reaction (\ref{eq:qqgluon}) and 
the soft limit of the phase space measure. The integration over the
gluon momentum can then be carried out analytically in $D$ dimensions. 
The two collinear regions are defined by 
$(\cos\theta_{qg}>(1-x_{\rm min})$ and $ x_g>x_{\rm min})$
and  $(\cos\theta_{qg}<(-1+x_{\rm min})$ and $ x_g>x_{\rm min})$, 
respectively, where
$\theta_{qg}$ is the angle between the gluon and the quark in the 
$q\bar{q}$ c.m. frame. In these regions we used the collinear 
approximations for both the squared matrix element and 
the phase space in $D$ dimensions.
Finally, the exact spin density matrix  for reaction (\ref{eq:qqgluon})
in four space-time dimensions was used in the resolved region, where
all necessary phase space integrations can be carried out numerically.
By construction, all four individual contributions depend 
logarithmically on the slicing parameter $x_{\rm min}$, but in the
sum only a residual linear dependence on $x_{\rm min}$ remains, which is due
to the approximations made in the soft and collinear regions.
By varying $x_{\rm min}$ between
$10^{-3}$ and $10^{-8}$ we checked
that for $x_{\rm min}\le 10^{-4}$ this residual dependence
is smaller than our numerical error (which is less than a permill
for all results discussed below). 
\par
After mass factorisation we are left with finite density matrices for the
$t\bar t$
and the $t\bar t$ + hard gluon final states. 
As a check of our calculation we first compute the total cross section for
$q\bar q\to t{\bar t} + X$ at NLO. 
If one identifies the $\overline{\rm{MS}}$ renormalisation scale $\mu$
with the mass factorisation scale $\mu_F$  and neglects all quark
masses except for $m_t$, then one can express the cross section in
 terms of dimensionless scaling functions \cite{Nason:1988}:
\begin{eqnarray}
\hat{\sigma}_{q\bar q}(\shat,m^2_t)&=&\frac{\alpha_s^2}{m^2_t}[
f^{(0)}_{q\bar q}(\eta) + 
4\pi\alpha_s(f^{(1)}_{q\bar q}(\eta)\nonumber \\  &+&
{\tilde f}^{(1)}_{q\bar q}(\eta) \ln(\mu^2/m^2_t))],
\label{eq:xsection}
\end{eqnarray}
where $\shat$ is the parton c.m. energy squared and $\eta = \shat/4m^2_t -1$.
We have compared our result for ${\sigma}_{q\bar q}$ as a function of
$\eta$
with those of \cite{Nason:1988,Beenakker:1991} and found perfect agreement.
\par
We now consider the following set of spin-correlation
observables:
\begin{equation}
{\cal O}_1=4\,\sp\cdot\sm,
\label{eq:sbasis}
\end{equation}
\begin{equation}
{\cal O}_2=4\,(\kh_1\cdot\sp)(\kh_2\cdot\sm),
\label{eq:hbasis}
\end{equation}
\begin{equation}
{\cal O}_3=4\,(\ph_1\cdot\sp)(\ph_1\cdot\sm),
\label{eq:pbasis}
\end{equation}
\begin{equation}
{\cal O}_4=4\,(\ph_2^*\cdot\sp)(\ph_1^{**}\cdot\sm),
\label{eq:ybasis}
\end{equation}
\begin{equation}
{\cal O}_5=4\,(\dhh_1\cdot\sp)(\dhh_2\cdot\sm),
\label{eq:obasis}
\end{equation}
where $\sp,\sm$ are the $t$ and $\bar t$ spin operators, respectively. 
The factor of 4 is conventional. With this normalization, 
the expectation value of ${\cal O}_1$ is equal to 1 at the Born level.
The expectation values of the  observables
(\ref{eq:hbasis}), (\ref{eq:pbasis}), (\ref{eq:ybasis}), 
and (\ref{eq:obasis}) determine the
correlation of different $t,\bar t$ spin projections.
Eq. (\ref{eq:hbasis}) corresponds to a 
correlation of the $t$ and $\bar t$ spins
in the helicity basis, while (\ref{eq:pbasis}) correlates the spins projected
along the beam line in the parton c.m.s. 
The `beam-line basis' used in (\ref{eq:ybasis}) was defined
 in \cite{Mahlon:1996} and refers to spin axes being parallel to the
antiquark direction
in the $t$ rest frame $\ph_2^*$ 
and to the quark direction in the
$\bar t$ rest frame  $\ph_1^{**}$, respectively. The spin axes
$\dhh_{1,2}$ in (\ref{eq:hbasis}) correspond to the so-called `optimal
basis' \cite{Parke:1996,Mahlon:1997} to be discussed below.
\par
For quark-antiquark
annihilation
it turns out that the spin
correlation (\ref{eq:pbasis}) \cite{Brandenburg:1996,Chang:1996}
and the correlation in the beam-line basis (\ref{eq:ybasis})
\cite{Mahlon:1996} are  stronger  than the correlation in the helicity basis.
A spin-quantization axis was constructed in \cite{Parke:1996,Mahlon:1997}
with respect to which the $t$ and $\bar t$
spins are 100$\%$ correlated to leading order in the
QCD coupling,  for all energies and scattering angles.
In terms of the structure functions of (\ref{eq:Rstruct}) this means
that the `optimal' spin axis $\dhh$ fulfills the condition
\begin{equation}
\hat{d}_i C_{ij} \hat{d}_j = A.
\label{eq:optcons}
\end{equation}
The existence of a solution of Eq. (\ref{eq:optcons}) is a special
property of the leading order spin density matrix for 
the reaction $q\bar{q}\to t\bar{t}$. One finds \cite{Parke:1996,Mahlon:1997}:
\begin{equation}
\dhh = \frac{-\ph_1+(1-\gamma_1)(\ph_1\cdot\kh_1)\kh_1}
{\sqrt{1-(\ph_1\cdot\kh_1)^2(1-\gamma_1^2)}}
, 
\end{equation}
where $\gamma_1=E_1/m_t$.
The construction of this axis 
explicitly uses the leading order result for the spin density matrix, and
different generalizations to higher orders are possible. We use in
(\ref{eq:obasis}) as spin axes:
\begin{eqnarray}
& &\dhh_1=\dhh, \nonumber \\
& & \dhh_2 = \frac{-\ph_1+(1-\gamma_2)(\ph_1\cdot\kh_2)\kh_2}{\sqrt{1-(\ph_1\cdot\kh_2)^2(1-\gamma_2^2)}}
, 
\end{eqnarray}
where  $\gamma_2=E_2/m_t$. For the 2 to 2 process $q\bar{q}\to t\bar{t}$,
$\dhh_2=\dhh_1=\dhh$. 
\par
The expectation value of a spin-correlation observable ${\cal O}$
at parton level can be written at next-to-leading order in analogy
 to (\ref{eq:xsection}) as follows:
\begin{eqnarray}
\langle {\cal O} \rangle_{q\bar q}  &=&
g^{(0)}_{q\bar q}(\eta) 
+ 4\pi\alpha_s(g^{(1)}_{q\bar q}(\eta)\nonumber \\&+&
{\tilde g}^{(1)}_{q\bar q}(\eta) \ln(\mu_F^2/m^2_t)).
\label{eq:expval}
\end{eqnarray}
Note that these quantities  depend explicitly 
on the factorization scale $\mu_F$, but only implicitly (through $\alpha_s$)
on the renormalization scale $\mu$. This is because a factor $\alpha_s^2$
drops out in the expectation values, and hence the Born result is of order
$\alpha_s^0$. 
\par
Our results for the functions 
$g^{(0)}_{q\bar q}(\eta),\ g^{(1)}_{q\bar q}(\eta)$ and
${\tilde g}^{(1)}_{q\bar q}(\eta)$ are shown for the five observables
(\ref{eq:sbasis})-(\ref{eq:obasis}) in Figs. 1-5. 
In each figure,
the dotted line is the Born result $g^{(0)}_{q\bar q}(\eta)$, the full
line shows the function $g^{(1)}_{q\bar q}(\eta)$, and the dashed
line is ${\tilde g}^{(1)}_{q\bar q}(\eta)$. A general feature of all 
results is that the QCD corrections are very small for values of
$\eta\lesssim 1$. For larger values of $\eta$, the functions 
$g^{(1)}_{q\bar q}(\eta)$ depart significantly from zero. Also, 
the functions ${\tilde g}^{(1)}_{q\bar q}(\eta)$ become nonzero, with
less dramatic growth as $\eta\to \infty$ and with an opposite sign
as compared to $g^{(1)}_{q\bar q}(\eta)$. The phenomenological 
implications of these features for spin correlations
at the Tevatron and the LHC will be studied in detail
in a future work. Here we merely note that the substantial QCD corrections
for large $\eta$ will be damped by the parton distribution functions,
which decrease rapidly with $\eta$. Moreover, at Tevatron energies values
of $\eta$ above $\sim 30$ are kinematically excluded.
\par
To summarize: We have computed the 
spin density matrices describing $t\bar{t}$ production 
by $q\bar{q}$ annihilation to order $\alpha_s^3$. Further we have evaluated
the scaling functions encoding the QCD corrections to 
spin correlations, using a number of different spin quantization axes.
This work provides a building block, which was missing so far,
towards a complete description
of the hadronic production of top quark pairs at NLO in
the strong coupling.
\begin{figure}
\unitlength1.0cm
\begin{center}
\begin{picture}(5.5,5.5)
\put(-1,-1){\psfig{figure=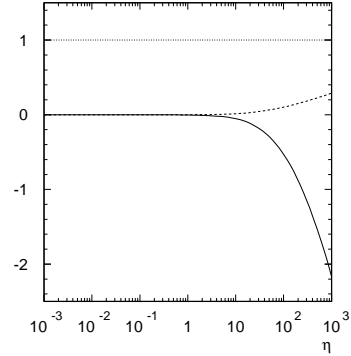,width=5.5cm,height=5.5cm}}
\end{picture}
\vskip -1.5cm
\caption{Dimensionless scaling functions $g^{(0)}_{q\bar q}(\eta)$ 
(dotted), $g^{(1)}_{q\bar q}(\eta)$ (full), and 
${\tilde g}^{(1)}_{q\bar q}(\eta)$ (dashed) that determine 
the expectation value $\langle {\cal O}_1 \rangle_{q\bar q}$.}\label{fig:obs1}
\end{center}
\end{figure}
\begin{figure}
\unitlength1.0cm
\begin{center}
\begin{picture}(5.5,5.5)
\put(-1,-1.){\psfig{figure=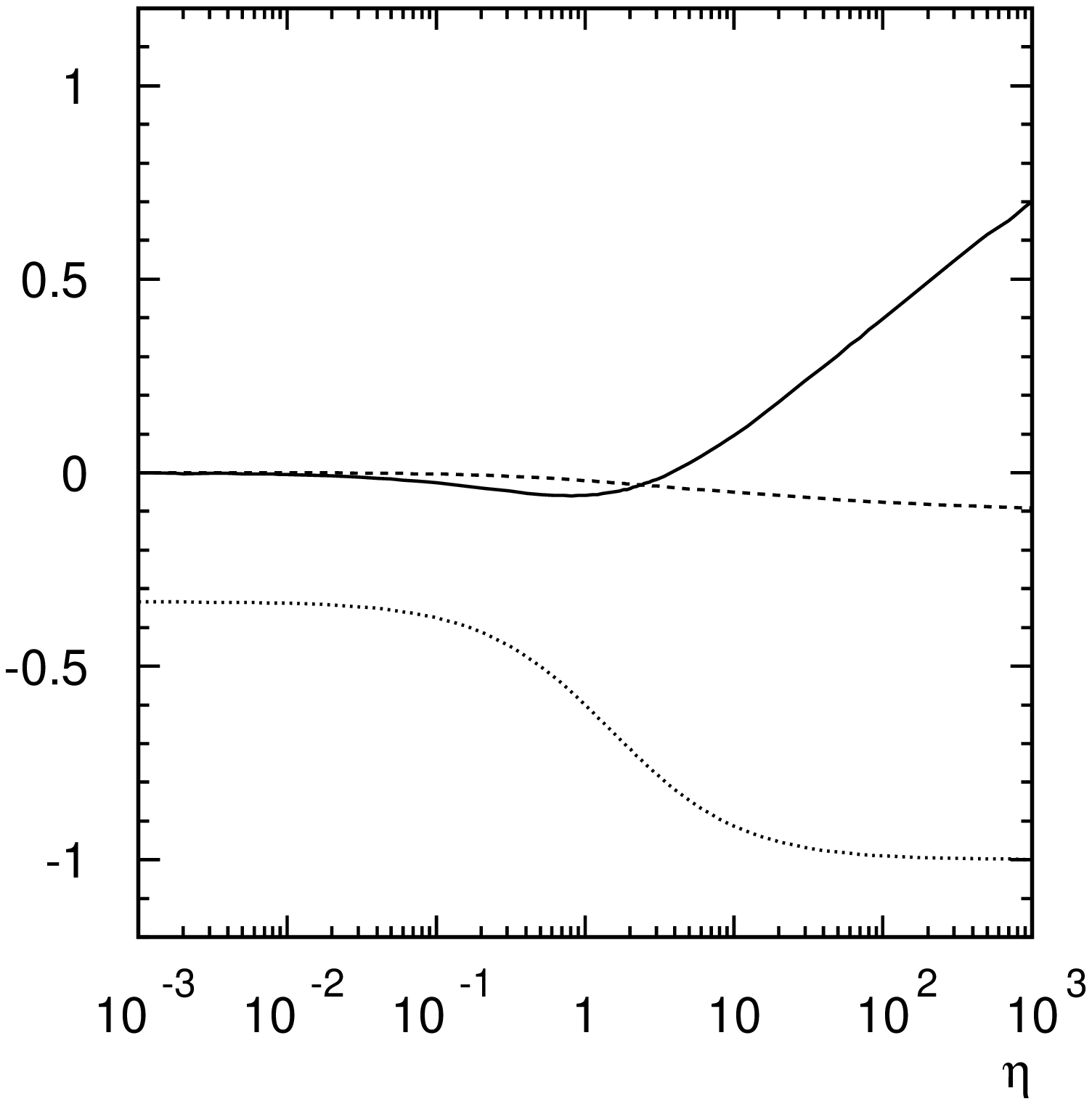,width=5.5cm,height=5.5cm}}
\end{picture}
\vskip -1.5cm
\caption{Same as Fig.1, but for $\langle {\cal O}_2 \rangle_{q\bar q}$.}
\label{fig:obs2}
\end{center}
\end{figure}
\begin{figure}
\unitlength1.0cm
\begin{center}
\begin{picture}(5.5,5.5)
\put(-1,-1){\psfig{figure=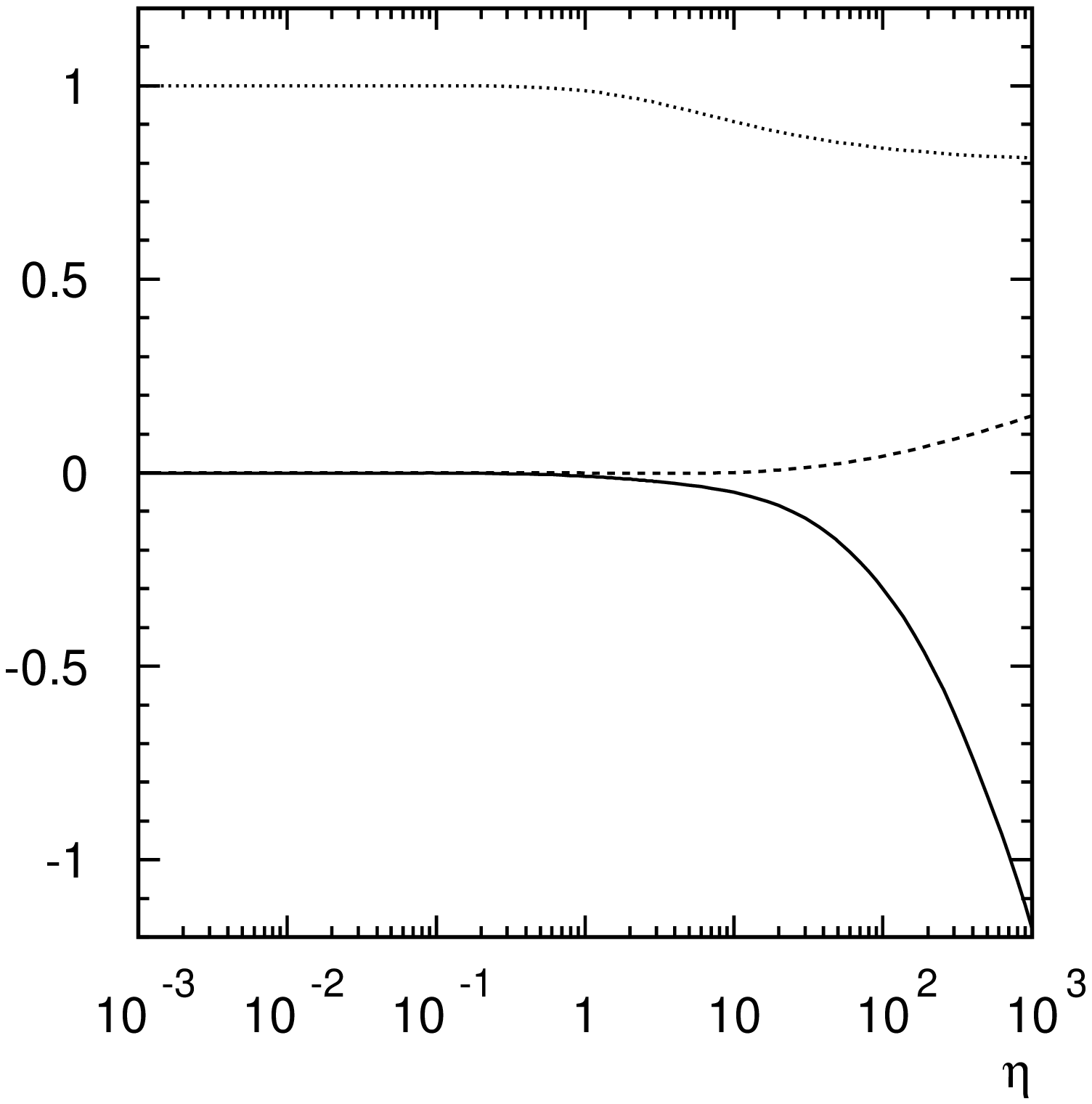,width=5.5cm,height=5.5cm}}
\end{picture}
\vskip -1.5cm
\caption{Same as Fig.1, but for $\langle {\cal O}_3 \rangle_{q\bar q}$.}
\label{fig:obs3}
\end{center}
\end{figure}
\newpage
\begin{figure}
\unitlength1.0cm
\begin{center}
\begin{picture}(5.5,5.5)
\put(-1,-1){\psfig{figure=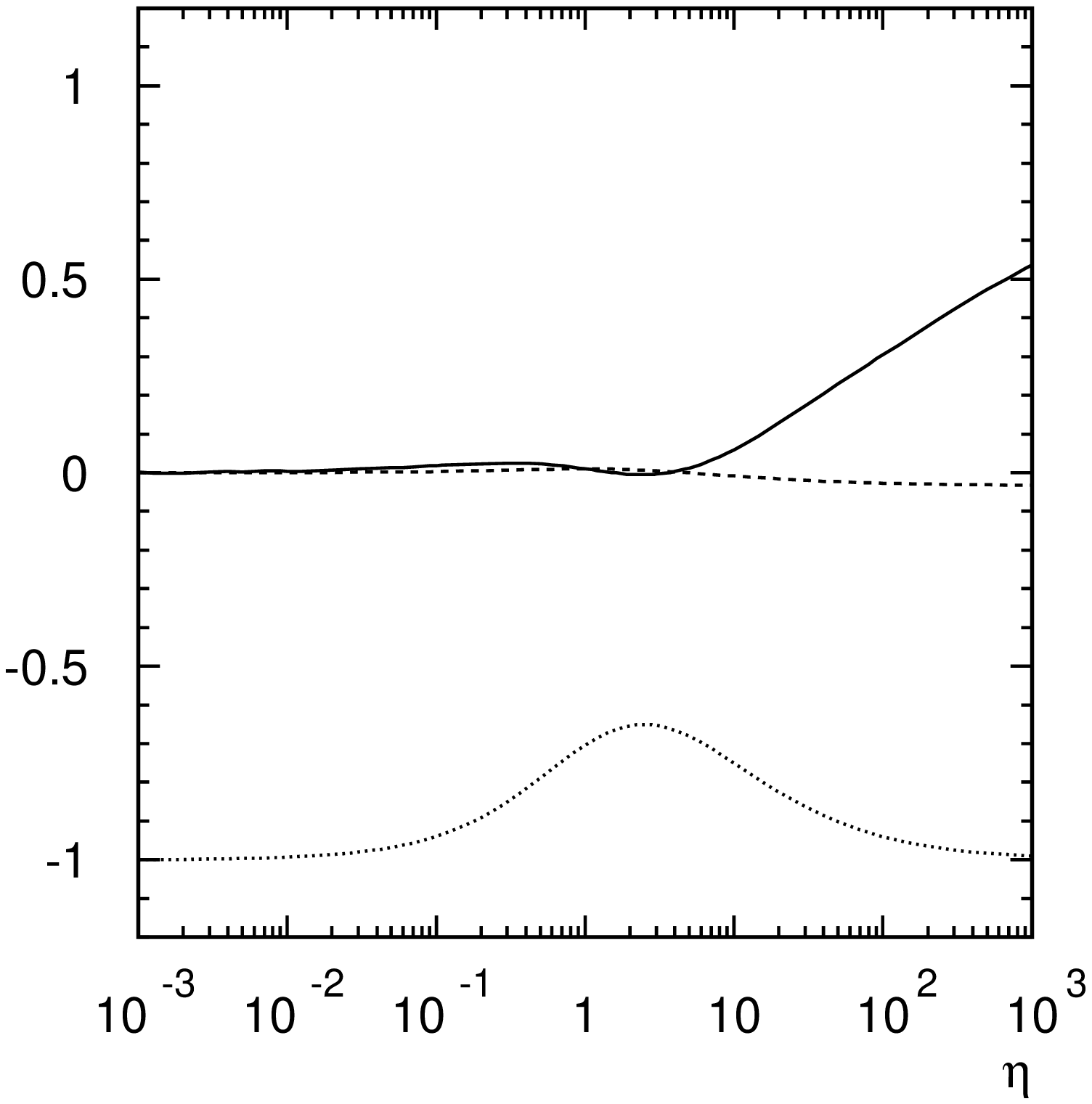,width=5.5cm,height=5.5cm}}
\end{picture}
\vskip -1.5cm
\caption{Same as Fig.1, but for $\langle {\cal O}_4 \rangle_{q\bar q}$.}
\label{fig:obs4}
\end{center}
\end{figure}
\begin{figure}
\unitlength1.0cm
\begin{center}
\begin{picture}(5.5,5.5)
\put(-1,-0.5){\psfig{figure=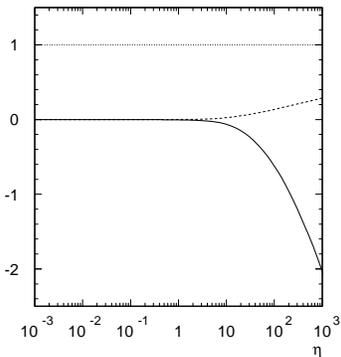,width=5.5cm,height=5.5cm}}
\end{picture}
\vskip -2cm
\caption{Same as Fig.1, but for $\langle {\cal O}_5 \rangle_{q\bar q}$.}
\label{fig:obs5}
\end{center}
\end{figure}

\subsubsection*{Discussion}
\noindent {\bf A.P. Contogouris,} Univ. of Athens 
\newline {\it I am worried not by the method of
your calculation, but by the importance of your numerical results. In 
particular the channel $gg\to t\bar{t}(g)$ can change them very much, and I am glad that you have already set up its calculation. Of course, $q\bar{q}\to
t\bar{t}(g)$ is required to be complete.}
\newline
\noindent {\bf A. Brandenburg}
\newline{\it I agree. Note that the plots show results at the parton level for
$q\bar{q}\to t\bar{t}(g)$.}

\begin{thebibliography}{99}

\bibitem{Kleiss:1988}
R.~Kleiss and W.~J.~Stirling,
Z.\ Phys.\  C\ 40 (1988) 419.

\bibitem{Stuart:1991}
R. G. Stuart, Phys. Lett. B 262 (1991) 113;
A. Aeppli, G. J. van Oldenborgh and D. Wyler, Nucl. Phys. B 428 (1994) 126.

\bibitem{Beenakker:1999}
W.~Beenakker, F. A. Berends and A. P. Chapovsky,
Phys.\ Lett.\ B\ 454 (1999) 129.

\bibitem{Nason:1988}
P.~Nason, S.~Dawson and R.~K.~Ellis,
Nucl.\ Phys.\  B\ 303 (1988) 607; Nucl.\ Phys.\  B\ 327  (1989) 49.

\bibitem{Beenakker:1991}
W.~Beenakker, W.~L.~van Neerven, R. Meng, G. A. Schuler and J.~Smith,
Nucl.\ Phys.\ B\ 351 (1991) 507.

\bibitem{Beenakker:1989}
W.~Beenakker, H.~Kuijf, W.~L.~van Neerven and J.~Smith,
Phys.\ Rev.\  D\ 40 (1989) 54.

\bibitem{Bernreuther:1996}
W.~Bernreuther, A.~Brandenburg and P.~Uwer,
Phys.\ Lett.\   B\ 368 (1996) 153.

\bibitem{Dharmaratna:1996}
W.~G.~Dharmaratna and G.~R.~Goldstein,
Phys.\ Rev.\  D\ 53 (1996) 1073.

\bibitem{Bernreuther:1994}
W.~Bernreuther and A.~Brandenburg,
Phys.\ Rev.\   D\ 49 (1994) 4481.


\bibitem{Brandenburg:1996}
A.~Brandenburg,
Phys.\ Lett.\   B\ 388 (1996) 626.


\bibitem{Barger:1989}
V.~Barger, J.~Ohnemus and R.~J.~Phillips,
Int.\ J.\ Mod.\ Phys.\  A\ 4 (1989) 617.

\bibitem{Stelzer:1996}
T.~Stelzer and S.~Willenbrock,
Phys.\ Lett.\   B\ 374 (1996) 169.


\bibitem{Mahlon:1996}
G.~Mahlon and S.~Parke,
Phys.\ Rev.\   D\ 53 (1996) 4886.

\bibitem{Mahlon:1997}
G.~Mahlon and S.~Parke,
Phys.\ Lett.\  B\ 411 (1997) 173.

\bibitem{Chang:1996}
D.~Chang, S.~Lee and A.~Sumarokov,
Phys.\ Rev.\ Lett.\  77 (1996) 1218.


\bibitem{D0:2000}
B. Abbott {\em et al.} (D0 Collaboration), 
Phys.Rev.Lett. 85 (2000) 256.

\bibitem{Top:2000}
M. Beneke et. al, 
``Top Quark Physics'' in:
Report of the ``1999 CERN Workshop on SM physics (and more) at the LHC'',
hep-ph/0003033.

\bibitem{Bernreuther:2000}
W. Bernreuther, A. Brandenburg and Z.G. Si,
Phys.Lett. B483 (2000) 99.

\bibitem{Czarnecki:1991}
A.~Czarnecki, M.~Jezabek and J.~H.~K\"uhn,
Nucl.\ Phys.\ B 351 (1991) 70.

\bibitem{Schmidt:1996}
C.~R.~Schmidt,
Phys.\ Rev.\   D 54 (1996) 3250.

\bibitem{Fischer:1999}
M.~Fischer, S.~Groote, J.~G.~K\"orner, M.~C.~Mauser and B.~Lampe,
Phys.\ Lett.\   B 451 (1999) 406.

\bibitem{Giele:1993}
W. T. Giele,  E. W. N. Glover and D. A. Kosower, Nucl.\ Phys.\ B\ 403
(1993) 633.

\bibitem{Parke:1996}
S. Parke and Y. Shadmi, Phys.\ Lett. B 387 (1996) 199.

\end{thebibliography}
\end{document}